\documentclass[aps,pra,reprint,showpacs,superscriptaddress,amsfonts,amsmath,amssymb]{revtex4-1}

\usepackage{lmodern}
\usepackage{graphicx}
\usepackage{xcolor}
\usepackage{xfrac}
\usepackage{dsfont}
\definecolor{blue}{rgb}{0.0,0.4,1.0}
\usepackage[colorlinks=true,citecolor=blue,linkcolor=blue,urlcolor=blue,pdfborder={0 0 0},breaklinks]{hyperref}

\newcommand{\ket}[1]{\ensuremath{\left|#1\right\rangle}}
\newcommand{\bra}[1]{\ensuremath{\left\langle#1\right|}}

\newcommand{\braketop}[3]{\left\langle#1\middle|#2\middle|#3\right\rangle}
\newcommand{\ketp}{\ket{{+}}}
\newcommand{\ketm}{\ket{{-}}}
\newcommand{\ketz}{\ket{{0}}}
\newcommand{\ketpm}{\ket{{\pm}}}
\newcommand{\keto}{\ket{1}}

\newcommand{\bop}{\ensuremath{b^{\vphantom\dagger}}}
\newcommand{\bopd}{\ensuremath{b^\dagger}}
\newcommand{\psiop}{\ensuremath{\psi^{\vphantom\dagger}}}
\newcommand{\psiopd}{\ensuremath{\psi^\dagger}}

\newcommand{\ef}[1]{\ensuremath{\operatorname{e}^{#1}}}
\newcommand{\bb}[1]{\left(#1\right)}
\newcommand{\hc}{\operatorname{\text{h.c.}}\!}
\newcommand{\absv}[1]{\left|#1\right|}

\renewcommand{\vec}[1]{\mathbf{#1}}
\newcommand{\veck}{\vec{k}}
\newcommand{\vecsigma}{\boldsymbol{\sigma}}
\renewcommand{\Re}{\mathop{\text{Re}}}
\renewcommand{\Im}{\mathop{\text{Im}}}

\newcommand{\figref}[1]{Fig.~\ref{fig:#1}}

\begin{document}

\title{Topological bands with Chern number \texorpdfstring{$C=2$}{C=2} by dipolar exchange interactions}

\author{David~Peter}
\email[Corresponding author: ]{peter@itp3.uni-stuttgart.de}
\affiliation{Institute for Theoretical Physics III, University of Stuttgart, 70569 Stuttgart, Germany}
\author{Norman~Y.~Yao}
\affiliation{Physics Department, Harvard University, Cambridge, Massachusetts 02138, USA}
\author{Nicolai~Lang}
\affiliation{Institute for Theoretical Physics III, University of Stuttgart, 70569 Stuttgart, Germany}
\author{Sebastian~D.~Huber}
\affiliation{Institute for Theoretical Physics, ETH Z\"urich, 8093 Z\"urich, Switzerland}
\author{Mikhail~D.~Lukin}
\affiliation{Physics Department, Harvard University, Cambridge, Massachusetts 02138, USA}
\author{Hans~Peter~B\"uchler}
\affiliation{Institute for Theoretical Physics III, University of Stuttgart, 70569 Stuttgart, Germany}

\date{\today}

\begin{abstract}
We demonstrate the realization of topological band structures by exploiting the intrinsic spin-orbit coupling of dipolar interactions in combination with broken time-reversal symmetry.
The system is based on polar molecules trapped in a deep optical lattice, where the dynamics of rotational excitations follows a hopping Hamiltonian which is determined by the dipolar exchange interactions.
We find topological bands with Chern number $C=2$ on the square lattice, while a very rich structure of different topological bands appears on the honeycomb lattice.
We show that the system is robust against missing molecules.
For certain parameters we obtain flat bands, providing a promising candidate for the realization of hard-core bosonic fractional Chern insulators.
%
%
%
%
%
\end{abstract}


\maketitle

\section{Introduction}
The quest for the realization of different topological states of matter marks one of the major challenges in quantum many-body physics.
A well established concept for the generation of two-dimensional topologically ordered states exhibiting anyonic excitations are flat bands characterized by a topological invariant in combination with strong interactions~\cite{Bergholtz2013,Parameswaran2013}.
The prime example is the fractional quantum Hall effect, where strong magnetic fields generate Landau levels~\cite{Nayak2008}.
Furthermore, lattice models without Landau levels have been proposed for the realization of topological bands~\cite{Haldane1988,Raghu2008,Wang2011,Neupert2011,Wang2012a,Grushin2012,Moller2009,Sun2010,Barkeshli2012,Wang2011a,Sterdyniak2013,Liu2012,Yao2013,Yang2012,Dauphin2012,Cooper2012,Cooper2013,Shi2013}.
Notably, spin-orbit coupling has emerged as an experimentally promising tool for band structures with topological invariants~\cite{Kane2005,Pesin2009,Qi2011,Hasan2010,Tang2011,Qiao2011}.
In this letter, we show that dipolar interactions, exhibiting intrinsic spin-orbit coupling, can be exploited for the realization of topological bands with cold polar molecules.

In cold gases experiments, the phenomenon that dipolar interactions exhibit spin-orbit coupling is at the heart of demagnetization cooling~\cite{Hensler2003,Fattori2006,Pasquiou2011,DePaz2013a},
and has been identified as the driving mechanism for the Einstein-de Haas effect in Bose-Einstein condensates~\cite{Kawaguchi2006} and the pattern formation in spinor condensates~\cite{Santos2006,Vengalattore2008,Kurn2013}.
Dipolar relaxation was proposed as a mechanism to reach the quantum Hall regime by the controlled insertion of orbital angular momentum~\cite{Peter2013}.
Recently, it has been pointed out that dipolar spin-orbit coupling can be observed in band structures realized with polar molecules~\cite{Syzranov2014}.
These ideas are motivated by the experimental success in cooling and trapping polar molecules in optical lattices~\cite{Ni2008b,Yan2013}.


\begin{figure}[t]
    \centering
    \includegraphics[width=.97\columnwidth]{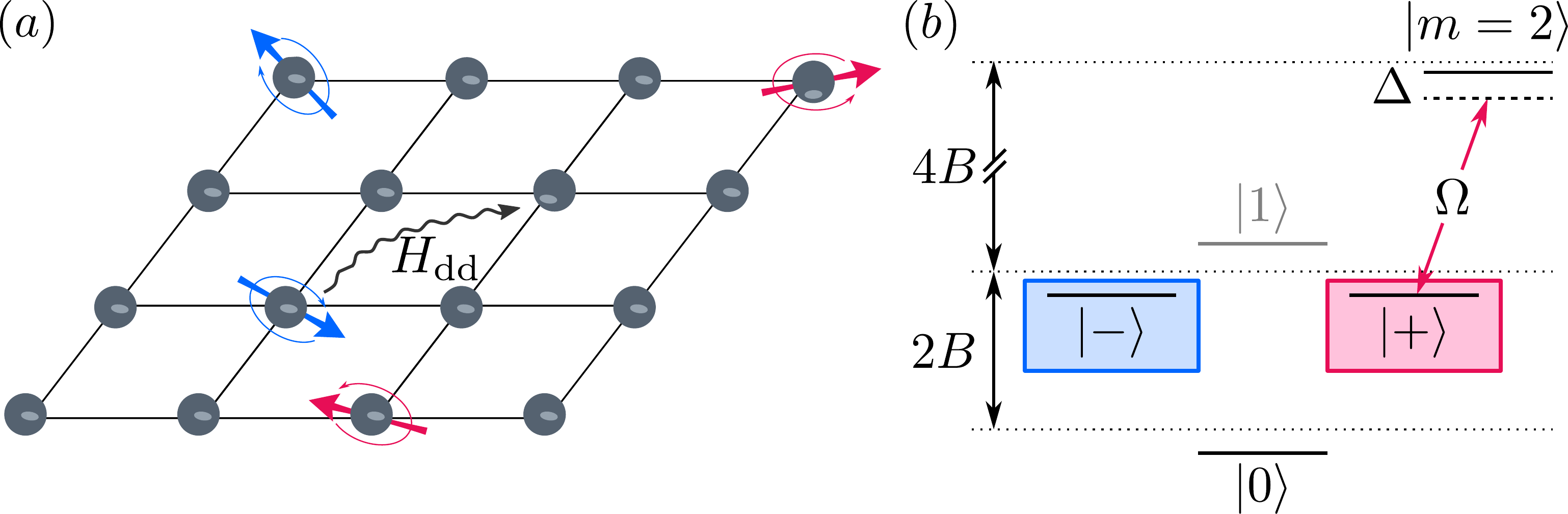}

    \caption{
        \label{fig:fig1}
        (a)~Setup: Each lattice site of a two-dimensional optical lattice is occupied by a single polar molecule.
        The molecules can be excited into two different rotational states.
        Dipole-dipole interactions induce long-range tunneling links for the excitations.
        (b)~Rotational level structure of each molecule with applied electric field and additional microwave field with Rabi frequency $\Omega$ and detuning $\Delta$.
    }
\end{figure}

Here we show that a system of polar molecules gives rise to topological band structures, exploiting the spin-orbit coupling of dipolar interactions in combination with a term that breaks time-reversal symmetry.
The main idea is based on polar molecules trapped in a two-dimensional deep optical lattice with quenched tunneling between the sites.
The relevant degree of freedom of the polar molecules is given by two different rotational excitations which can be transferred between different lattice sites due to the dipolar exchange interaction.
We demonstrate that the band structure for such an excitation is characterized by a Chern number which depends on the underlying lattice structure.
In particular, we find that the system on a square lattice gives rise to Chern number $C=2$, while a rich phase diagram appears on the honeycomb lattice. 
Ideally, the setup is initialized with one polar molecule per lattice site, but we demonstrate that the topological properties are robust, even if nearly half of the molecules are randomly removed.
In contrast to non-interacting fermions, free bosons cannot form a topological insulator. However,
the bosonic excitations in our system are subject to a hard-core constraint. Such a setup in combination with flat bands is then expected to give rise to a fractional Chern insulator at $2/3$ filling in $C=2$ topological bands~\cite{Moller2009,Wang2012a,Sterdyniak2015,Yao2015}.

The main advantages of our realization, using the spin-orbit coupling present in dipolar interactions, are its robustness and the low experimental requirements, while many alternative theoretical proposals with cold gases require strong spatially inhomogeneous laser fields with variations on the scale of one lattice constant~\cite{Liu2010,Stanescu2010,Goldman2013,Li2008,Yao2012,Yao2013,Goldman2013,Jaksch2003}; by using such ideas in combination with dipolar exchange interactions, it is also possible to engineer flat $C=2$ bands~\cite{Yao2015}.
We point out that our proposal can also be applied to Rydberg atoms in similar setups~\cite{Barredo2014,Piotrowicz2013,Nogrette2014}.

\section{Setup}
We consider a two-dimensional system of ultracold polar molecules in a deep optical lattice with one molecule pinned at each lattice site, as shown in~\figref{fig1}a.
The remaining degree of freedom is given by the internal rotational excitations of the molecules with the Hamiltonian
\begin{align}
    H^{\text{rot}}_i &= B \vec{J}_i^2 - \vec{d}_i\cdot\vec{E}\,.
\end{align}
Here, $B$ is the rotational splitting, $\vec{J}_i$ is the angular momentum of the $i$th molecule and $\vec{d}_i$ is its dipole moment which is coupled to the applied static and microwave electric fields $\vec{E} = \vec{E}_{\text{s}} + \vec{E}_{\text{ac}}(t)$.
In the absence of external fields, the eigenstates $\ket{J,m}$ of $H^{\text{rot}}_{i}$ are conveniently labeled by the total angular momentum $J$ and its projection $m$.
Applying a static electric field mixes states with different $J$.
The projection $m$, however, can still be used to characterize the states.
In the following, we focus on the lowest state $\ketz$ with $m=0$ and the two degenerate excited states $\ketpm$ with $m = \pm 1$, see~\figref{fig1}b.
The first excited $m=0$ state, called $\keto$, will be used later.



The full system, including pairwise dipole-dipole interactions between the polar molecules, is described by $H=\sum_i H^{\text{rot}}_i + \frac{1}{2}\sum_{i\ne j}H^{\text{dd}}_{ij}$.
In the two-dimensional setup with the electric field perpendicular to the lattice, the interaction can be expressed as
\begin{align}
    H^{\text{dd}}_{ij}=\frac{\kappa}{|\vec{R}_{ij}|^3} &\Big[d^0_i d^0_j + \frac{1}{2}\big(d^+_i d^-_j + d^-_i d^+_j) \nonumber \\
                                                       &\quad - \frac{3}{2}\big(d^-_i d^-_j \ef{2i\phi_{ij}} + d^+_i d^+_j \ef{-2i\phi_{ij}}\big)\Big] \label{eq:ddint}
\end{align}
with $\kappa=1/4\pi\epsilon_0$.
Here, $\phi_{ij}$ denotes the in-plane polar angle of the vector $\vec{R}_{ij} \equiv |\vec{R}_{ij}| \cdot (\cos \phi_{ij}, \sin \phi_{ij})^t$ which connects the two molecules at lattice sites $i$ and $j$, and the operators $d^0=d^z$ and $d^\pm=\mp (d^x\pm i d^y)/\sqrt{2}$ are the spherical components of the dipole operator.
The intrinsic spin-orbit coupling is visible in the second line in Eq.~\eqref{eq:ddint}, where a change in internal angular momentum by $\pm 2$ is associated with a change in orbital angular momentum encoded in the phase factor $e^{\mp 2i \phi_{ij}}$.

For molecules with a permanent dipole moment $d$ in an optical lattice with spacing $a$, the characteristic interaction energy $V=\kappa d^2/a^3$ is much weaker than the rotational splitting $B$.
For strong electric fields, the energy separation between the states $\ketpm_i$ and $\keto_i$ is also much larger than the interaction energy.
Then the number of $\ketpm$ excitations is conserved.
This allows us to map the Hamiltonian to a bosonic model: The lowest energy state with all molecules in the $\ketz$ state is the vacuum state, while excitations of a polar molecule into the state $\ket\pm_i$ are described by hard-core boson operators $\smash{\bopd_{i,\pm} =\ket{\pm}_i\!\bra{0}_i }$.
Note that these effective bosonic particles have a spin angular momentum of $m =\pm 1$.

A crucial aspect for the generation of topological bands with a nonzero Chern number is the breaking of time-reversal symmetry.
In our setup, this is achieved by coupling the state $|+\rangle_{i}$ to the rotational state $\ket{m=2}_{i}$ with an off-resonant microwave field \footnote{The coupling of the $\ketm$ state to the third $m=0$ state can be neglected due to a large detuning from the difference in Stark shifts between $m=0$ and $m=2$}, see \figref{fig1}b.
This coupling lifts the degeneracy between the two excitations $|\pm\rangle_{i}$ and provides an energy splitting denoted by $2 \mu$.


\section{Topological band structure}
The dipole-dipole interaction gives rise to an effective hopping Hamiltonian for the bosonic particles due to the dipolar exchange terms:
$d^+_id^-_j$, for example, leads to a (long-range) tunneling $\bopd_{i,+}\bop_{j,+}$ for the ${+}$-bosons while the term $d^-_i d^-_j \ef{2i\phi_{i j}}$ generates spin-flip tunneling processes $\bopd_{i,-}\bop_{j,+} \ef{2i\phi_{ij}}$ with a phase that depends on the direction of tunneling.
For the study of the single particle band structure we can drop the term proportional to $d^0 d^0$ which describes a static dipolar interaction between the bosons.
The interaction Hamiltonian reduces to
\begin{align}
    H^{\text{dd}} = \sum_{i\ne j}
    \frac{a^3}{|\vec{R}_{ij}|^3}\;
    \psiopd_i\!
    \begingroup
        \renewcommand*{\arraystretch}{1.2}
        \begin{pmatrix}
            -t^+ & w \ef{-2i\phi_{ij}} \\
            w \ef{2i\phi_{ij}} & -t^-
        \end{pmatrix}
    \endgroup\!
    \psiop_j \,,
    \label{eq:hrealspace}
\end{align}
where we use the spinor notation $\psiopd_j = \big( \bopd_{j,+} , \bopd_{j,-} \big)$.
The energy scale of the hopping rates $t^+$, $t^-$, and $w$ is given by $V$. The precise form depends on the microscopic parameters and is detailed in the appendix.
Note that $t^{+} = t^{-}$ without the applied microwave.
In momentum space with $\psiop_{\veck}=\frac{1}{\sqrt{N_s}}\sum_{j} \psiop_{j}\ef{i\veck\vec{R}_j}$, including the internal energy $H_{i}^{\text{rot}}$ of the excitations $|\pm\rangle_{i}$, the Hamiltonian can be rewritten as
\begin{align}
    H &= \sum_{\veck} \psiopd_{\veck} \big({n^{0\vphantom\dagger}_{\veck} \: \mathds{1} + \vec{n}^{\vphantom \dagger}_{\veck}\cdot\vecsigma}\big) \psiop_{\veck}
    \label{eq:hkspace}
\end{align}
where it is useful to express the traceless part of the Hamiltonian
as the product of a three dimensional real vector $\vec{n}_{\veck}$ and the vector of Pauli matrices $\vecsigma$ \cite{Hasan2010,Bernevig2013}.
The real vector characterizes the spin-orbit coupling terms and takes the form
\begin{align}
    \vec{n}^{\vphantom 0}_{\veck} = \begin{pmatrix}
        w \Re \epsilon^2_{\veck} \\
        w \Im \epsilon^2_{\veck} \\
        \mu + t\, \epsilon^0_{\veck}
    \end{pmatrix}
\end{align}
with $t = (t^--t^+)/2 > 0$.
The spin-independent hopping is determined by $n^0_{\veck} = -\bar{t}\, \epsilon^0_{\veck}$ with ${\bar t} = (t^{+}+t^{-})/2$.
We have introduced the dipolar dispersion relation~\cite{Peter2012b,Syzranov2014}
\begin{align}
    \epsilon^m_{\veck} = \sum_{j\ne 0} \frac{a^3}{|\vec{R}_j|^3}\ef{i\veck \vec{R}_j + i m \phi_{j}}.
\end{align}
The precise determination of this function can be achieved by an Ewald summation technique providing a non-analytic low momentum behavior
$\epsilon^0_{\veck} \approx \epsilon^0_{\Gamma} - 2\pi |\veck|a$ and
$\epsilon^2_{\veck} \approx -\frac{2\pi}{3} |\veck|a \ef{2i \varphi}$.
Here, $\epsilon^0_{\Gamma} \approx 9.03$ and $\varphi$ is defined by $\hat{\veck}=(\cos \varphi, \sin \varphi)^t$.

\begin{figure}[b]
    \centering
    \includegraphics[width=.48\columnwidth]{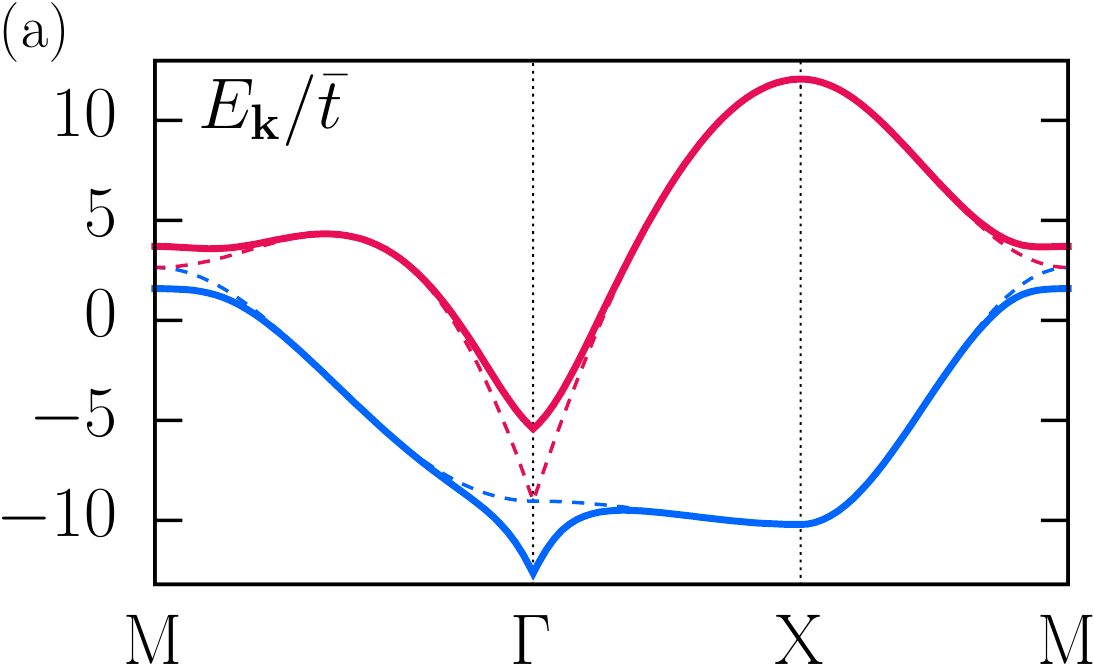}
    \includegraphics[width=.48\columnwidth]{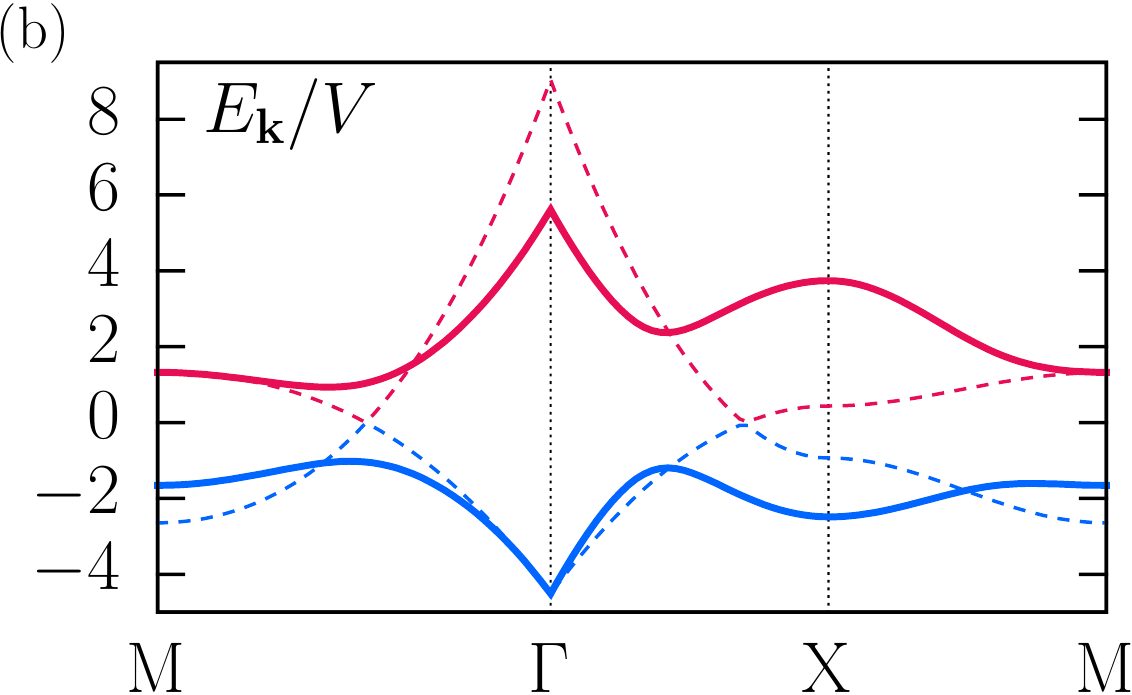}

    \caption{
        \label{fig:fig2}
        (a) Dispersion relation for the $\ketp$ and $\ketm$ states on the square lattice.
        The dashed line shows the time-reversal invariant point $t=\mu=0$ with band touching at the $\Gamma$ and $\text{M}$ point.
        The solid line shows the gapped topological bands in the time-reversal-broken system for $w/\bar{t}=3, \mu=0$ and $t/\bar{t}=0.4$.
        (b) Dispersion relation for the $\ketp$ and $\keto$ states for electric field angles $\Theta_0=0$ (dashed) and $\Theta_0=\pi/4$ (solid), respectively.
        The latter has a lower band with flatness $f\approx 1$.
    }
\end{figure}

In the presence of time-reversal symmetry, represented by $\mathcal{T}=\sigma_x \mathcal{K}$ with $\mathcal{K}$ being complex conjugation, the system reduces to the one discussed in ref.~\cite{Syzranov2014}.
At the $\mathcal{T}$-invariant point, i.e. $t=\mu=0$, the two energy bands of the system exhibit a band touching at the high-symmetry points $\Gamma=(0,0)$ and $\text{M}=(\pi/a, \pi/a)$ where $\epsilon^2_\veck$ vanishes, see \figref{fig2}a.
The touching at the $\Gamma$ point is linear due to the low-momentum behavior of $\epsilon^m_\veck$.
Note that each of the touching points splits into two Dirac points if the square lattice is stretched into a rectangular lattice.

Breaking of time-reversal symmetry by the microwave field leads to an opening of a gap between the two bands.
The dispersion relation is given by
\begin{align}
    E_\pm(\veck)=- \bar{t}\, \epsilon^0_\veck \pm \sqrt{w^2 \big|\epsilon^2_{\veck}\big|^2+\big(\mu+t\, \epsilon^0_{\veck}\big)^2}
\end{align}
and shown in \figref{fig2}a.
It is gapped whenever the vector $\vec{n}_{\veck}\ne 0$.
The first two components can only vanish at the $\Gamma$ or $\text{M}$ point.
Consequently, the gap closes iff the third component is zero at one of these two points, that is for
\begin{align} 
    \mu/t &= -\epsilon^0_{\Gamma} \approx -9.03, \nonumber\\
    \mu/t &= -\epsilon^0_\text{M} = \Big(1-1/\sqrt{2}\Big)\epsilon_{\Gamma}^0 \approx +2.65.
\end{align}
In the gapped system, the Chern number~\cite{Hasan2010,Qi2011} can be calculated as the winding number of the normalized vector $\hat{\vec{n}}_\veck = \vec{n}_\veck/|\vec{n}_\veck|$ via
\footnote{We remark that we need to truncate the summation in the expression for $\epsilon^m_{\veck}$ to perform the calculation of the Chern number.
We can check, however, that the remaining terms are not strong enough to close a gap.
Conversely, note that the cutoff radius has to be larger than $\sqrt{2}a$, as the next-to-nearest neighbor terms are crucial for the $C=2$ phase and may not be neglected.}
\begin{align}
    C &= \frac{1}{4\pi}\int_{\text{BZ}}\!\mathrm{d}{^2\veck}\, ( \partial_{k_x} \hat{\vec{n}}_{\veck} \times \partial_{k_y} \hat{\vec{n}}_{\veck} ) \cdot \hat{\vec{n}}_{\veck} \, . \label{eq:chern}
\end{align}
We find that the Chern number of the lower band is $C=2$ for $-\epsilon^0_{\Gamma} < \mu/t < -\epsilon^0_\text{M}$, and zero outside this range.
Note that the non-trivial topology solely results from dipolar spin-orbit coupling and time-reversal symmetry breaking.

\begin{figure}[t]
    \centering
    \includegraphics[width=.63\columnwidth]{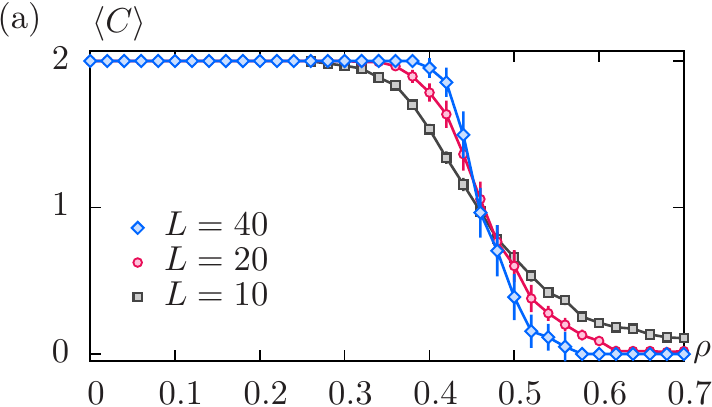}
    \hspace{0.1cm}
    \includegraphics[width=.32\columnwidth]{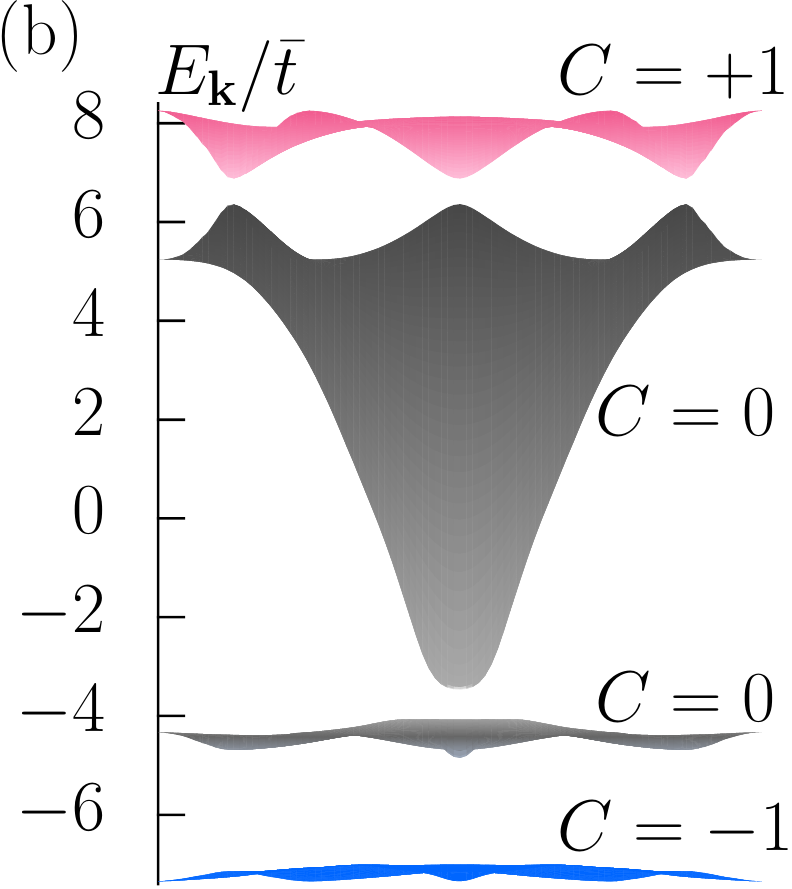}
    \includegraphics[width=\columnwidth]{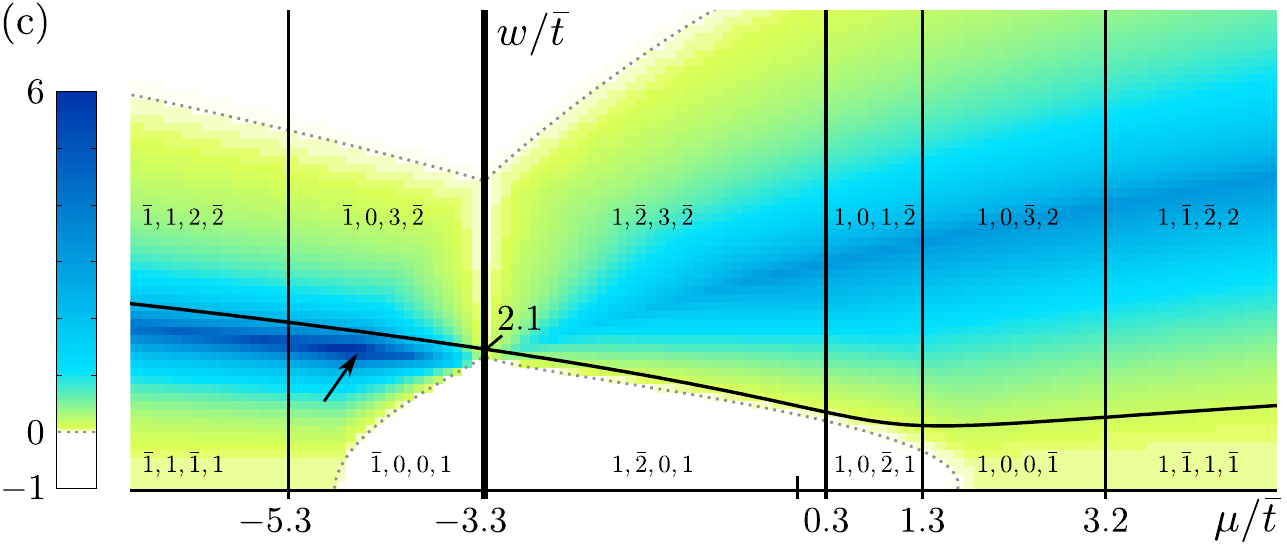}

    \caption{
        \label{fig:fig3}
        (a)~Sample-averaged Chern number $\langle C \rangle$ in the disordered system for increasing density $\rho$ of defects.
        A single realization either yields $C=2$ or $C=0$.
        Bars indicate two standard errors.
        The results are shown for square lattices of size $L \times L$ with $L=10, 20, 40$.
        The long-range tunneling stabilizes the topological phase for defect densities $\rho \lesssim 0.45$.
        (b)~Two-dimensional projection of the dispersion relation in the honeycomb lattice for $t/\bar{t}=0.54, w/\bar{t}=1.97$ and $\mu/\bar{t}=-4.54$.
        The lowest band has a flatness ratio of $f\approx 6.4$ and a Chern number of $C=-1$.
        (c)~Topological phase diagram in the honeycomb lattice for $t/\bar{t}=0.54$.
        The labels give the Chern numbers of the four bands (bar indicates negative number) from bottom to top while the solid lines correspond to touching points between two bands.
        The color indicates the flatness $f$ of the lowest band.
        The arrow shows the parameters of the flat-band model in~(b).
    }
\end{figure}

The challenge is to find a specific setup that optimizes the flatness of the topological bands.
This can be achieved either by focusing on different lattice structures (see honeycomb lattice below and \figref{fig3}b) or by an alternative choice for the two excitations.
The latter is less intuitive when trying to understand the spin-orbit coupling, but gives rise to significantly flatter bands: Instead of considering $\ketp$ and $\ketm$, we choose a model including the $\ketp$ and $\keto$ states.
This is possible for weak electric fields, if the $\ketm$ state is shifted by a microwave field, or by exploiting the coupling between the nuclear spins of the polar molecules and the rotational degree of freedom~\cite{Ospelkaus2010,Yan2013}.
This model intrinsically breaks time-reversal symmetry and has the advantage that the $\ketp$ and $\keto$ states have different signs for the tunneling strength, making the $\mathcal{T}$-breaking parameter $t=(t^+-t^1)/2$ large compared to $\bar{t}=(t^++t^1)/2$. 
For an electric field direction perpendicular to the lattice, this system is gapless.
Opening the gap is achieved by rotating the electric field away from the $z$-axis by an angle $\Theta$.
The dispersion relation for $\Theta=0$ and $\pi/4$ is shown in~\figref{fig2}b.
The lower band has a flatness ratio of $f = \text{bandgap}/\text{bandwidth} \approx 1$.

Topological band structures are classified by considering equivalence classes of models that can be continuously deformed into each other without closing the energy gap~\cite{Hasan2010}.
Using this idea, we can demonstrate that our model with $C=2$ is adiabatically equivalent to a system of two uncoupled copies of a $C=1$ layer (see appendix for details).
The resulting single layer model can be described by a staggered flux pattern and is reminiscent of the famous Haldane model~\cite{Haldane1988}, adapted to the square lattice~\cite{Goldman2013,Li2008,Liu2010,Liu2011,Stanescu2010,Wang2011,Wang2014,Yao2012,Yao2013}.
It is rather remarkable that uniform dipole-dipole interactions give rise to a model usually requiring strong modulations on the order of the lattice spacing.

\section{Influence of disorder}
An experimental initialization with a perfectly uniform filling of one molecule per site is challenging.
Consequently, we analyze the stability of the topological band structure for random samples with a nonzero probability $\rho$ for an empty lattice site.
The determination of the Chern number for the disordered system follows ideas from refs.~\cite{Niu1985,Avron1985}.
We start with a finite geometry of $L \times L$ lattice sites and twisted boundary conditions
$\psi(x + L, y) = \ef{i\theta_x}\psi(x, y)$ and $\psi(x, y + L) = \ef{i\theta_y}\psi(x, y)$ for the single particle wave function.
Next, we randomly remove $\rho L^2$ lattice sites (dipoles).
We are interested in the Chern number of the lower `band', composed of the lowest $N_l=L^2 (1-\rho)$ states (there are $2N_l$ states in total).
To this end, we pretend to have a free fermionic system at half filling whose many-body ground state $\Psi=\Psi(\theta_x, \theta_y)$ is given by the Slater determinant of the lowest $N_l$ states.
Then, the Chern number can be calculated as
\begin{align}
    C &= \frac{1}{2\pi}\iint\!\mathrm{d}\theta_x \mathrm{d}\theta_y \, F(\theta_x, \theta_y), \label{eq:manybodychern}
\end{align}
where $F(\theta_x,\theta_y)=\smash{\Im\!\big({\big\langle\frac{\partial \Psi}{\partial \theta_y}\big|\frac{\partial \Psi}{\partial \theta_x}\big\rangle - \big\langle\frac{\partial \Psi}{\partial \theta_x}\big|\frac{\partial \Psi}{\partial \theta_y}\big\rangle}\big)}$ is the many-body Berry curvature depending on the boundary condition twists.
Note that Eq.~\eqref{eq:manybodychern} reduces to~Eq.~\eqref{eq:chern} in the translationally invariant case.
For the numerical computations, we use a discretized version~\cite{Fukui2005}.
The results for the disordered system are summarized in~\figref{fig3}a.
We find that the long-range tunneling stabilizes the topological phase for defect densities $\rho\lesssim 0.45$.

\section{Honeycomb lattice}
Returning to the simple setup in \figref{fig1}b, the influence of the lattice geometry can be exemplified by going to the honeycomb lattice.
Due to the two distinct sublattices, we generally obtain four bands in the presence of broken time-reversal symmetry.
Depending on the microscopic parameters, the bands exhibit a rich topological structure, characterized by their Chern numbers.
Note that the Chern numbers are calculated with a numerical method similar to the one for the disordered system.
In \figref{fig3}c, we show a two-dimensional cut through the topological phase diagram, spanned by the parameters $t/\bar{t}, w/\bar{t}$ and $\mu/\bar{t}$.
We find a multitude of different topological phases with large areas of flatness $f > 0$ for the lowest band.
A flatness $f < 0$ indicates that the maximum of the lowest band is higher than the minimum of the second band.
In contrast to the square lattice, an energy splitting $\mu \ne 0$ is sufficient for a nonzero Chern number; $t\ne 0$ is not necessarily needed.
\figref{fig3}b shows the dispersion relation with a lowest band of flatness $f\approx 6.4$ and a Chern number $C=-1$.


\begin{figure}[t]
    \centering
    \includegraphics[width=.504\columnwidth]{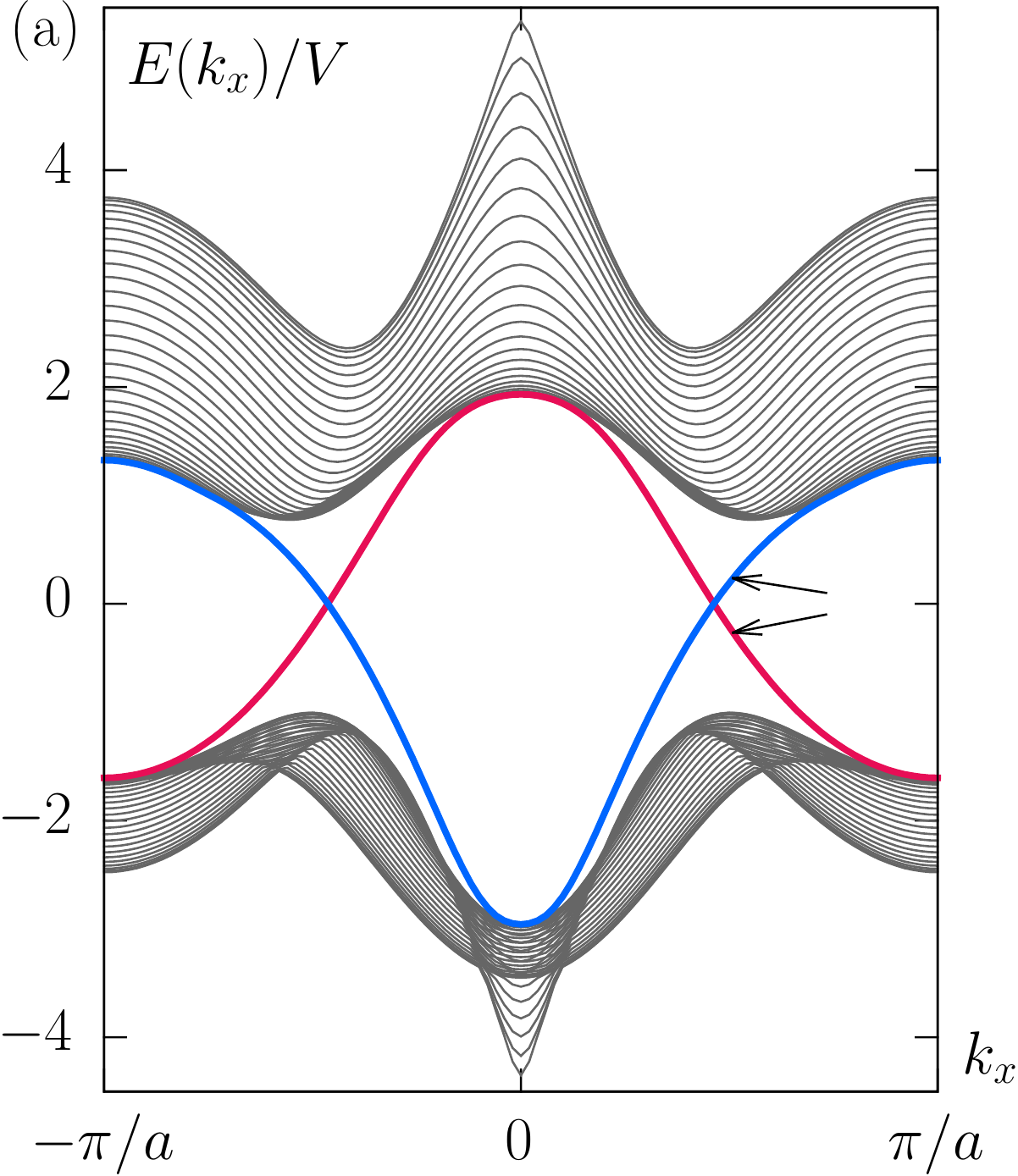}
    \begin{minipage}[b]{.482\columnwidth}
        \includegraphics[width=.95\columnwidth]{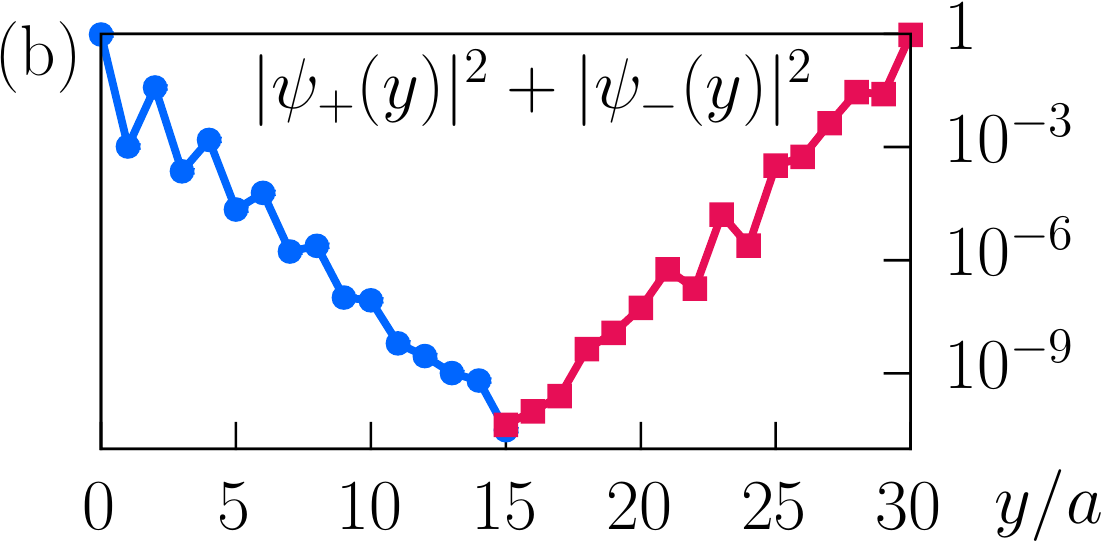} \\
        \vspace{2mm}
        \includegraphics[width=.95\columnwidth]{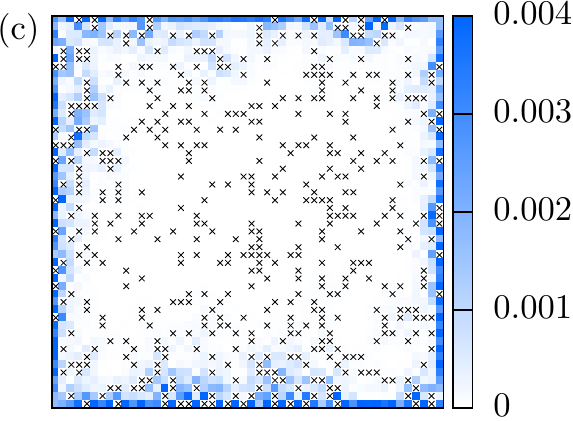}
    \end{minipage}

    \caption{
        \label{fig:fig4}
        (a)~Dispersion relation for the $\ketp$ and $\keto$ states on a cylindrical square lattice geometry with infinite extent in $x$ direction and $31$ sites in $y$ direction.
        Four edge states cross the bandgap in the $C=2$ phase (two for each edge).
        (b)~Exponentially decaying amplitude of the two edge states in logarithmic scale, corresponding to the points shown in the spectrum.
        (c)~Edge state amplitude $|\psi_+(x,y)|^2 + |\psi_-(x,y)|^2$ on a finite $50 \times 50$ square lattice with a filling fraction of $0.8$ (missing sites are indicated by crosses).
    }
\end{figure}
\section{Detection and outlook}
One way to detect the topological band structure experimentally is to create a local excitation close to the edge of the system.
In \figref{fig4} we show the edge states in the $C=2$ phase on the square lattice.
The states are exponentially localized on the boundary of the system and the propagation of a single excitation along the edge can be used as an indication of the topological nature of the bands~\cite{Hafezi2013}.
In \figref{fig4}c, we show the robustness of the edge states against missing molecules.
The edge state is also visible in a spectroscopic analysis, as a single mode between the broad continuum of the two bands (see~\figref{fig4}a).

Finally, the most spectacular evidence of the topological nature would be the appearance of fractional Chern insulators in the interacting many-body system at a fixed density of excitations.
In our system, the hard-core constraint naturally provides a strong on-site interaction for the bosons.
In addition, the remaining static dipolar interactions are a tunable knob to control the interaction strength.
The most promising candidate for a hard-core bosonic fractional Chern insulator in a band with $C=2$ appears for a filling of $\nu = 2/3$, as suggested by numerical calculations~\cite{Moller2009,Wang2012a,Yao2015}, in agreement with the general classification scheme for interacting bosonic topological phases~\cite{Lu2012a,Chen2013}.

\begin{acknowledgments}
We acknowledge the support of the Center for Integrated Quantum Science and Technology (IQST),
the Deutsche For\-schungs\-ge\-mein\-schaft (DFG) within the SFB/TRR~21
and the Swiss National Science Foundation.
\end{acknowledgments}

\begin{appendix}

\section{Microscopic form of the parameters}
As described in the main text, in the presence of the static electric field, we denote the lowest rotational states having $m=0, \pm 1$ by $\ketz$ and $\ketpm$, respectively.
In addition, let $\ket{m=2}$ be the lowest $m=2$ state.
A microwave with Rabi frequency $\Omega \equiv 2 E_\text{ac} \absv{\braketop{m=2}{d^+}{{+}}}$ and detuning $\Delta$ couples the states $\ketp$ and $\ket{m=2}$.
For a large detuning $\Delta \gg \Omega, V$, the number of $\ketp$ (and $\ketm$) excitations is conserved.
In the rotating frame, within rotating wave approximation, the AC-dressed $\ketp$ state is given by
\begin{align}
    \ketp_\text{ac} &= \bb{1-\epsilon^2/2}\ketp - \epsilon \ket{m=2}
\end{align}
up to second order in $\epsilon=\Omega/2\Delta$.
For sufficiently strong electric fields, the states $\ketz$ and $\ketm$ are essentially unaffected by the microwave.

To derive the parameters of the hopping Hamiltonian~(3) in the main text, we introduce the vacuum state $\ket{\text{vac}}=\prod_k \ketz_k$ and the single particle states $\ket{\Psi_{i,\pm}}=\bopd_{i,\pm} \ket{\text{vac}}$.
Then, the hopping amplitudes are given by
\begin{align}
    t^{\alpha\beta}_{ij} &= \braketop{\Psi_{i,\alpha}}{H^\text{dd}}{\Psi_{j,\beta}}.
\end{align}
We set the spin-conserving term $t^{\pm\pm}_{ij}=-t^\pm \cdot a^3/|\vec{R}_{ij}|^3$ and the spin-flip tunneling $t^{-+}_{ij}=w \ef{2i \phi_{ij}} \cdot a^3/|\vec{R}_{ij}|^3$ to get the final expressions for the nearest-neighbor tunneling rates
\begin{align}
    t^+ &= \frac{\kappa d_{1}^2}{2a^3} (1 - \epsilon^2), \nonumber \\
    t^- &= \frac{\kappa d_{1}^2}{2a^3}, \nonumber \\
    w &= \frac{3 \kappa d_{1}^2}{2a^3} (1-\epsilon^2/2)
\end{align}
with the transition dipole element $d_{1}=\absv{\braketop{-}{d^{-\vphantom{0}}}{{0}}}$.
The evaluation of the dipole matrix elements for finite electric fields is straightforward and has been described in detail~\cite{Micheli2007}.

In the absence of other techniques to shift the energy, the expression for the offset between the $\ketp$ and $\ketm$ state is given by the AC Stark shift $\Omega^2/4\Delta\equiv 2\mu$.

\emph{Model with $\ketp$ and $\keto$ state:}
Using the $\ketp$ and the $\keto$ state (first excited $m=0$ state), the microwave field is no longer necessary, as the model intrinsically breaks time-reversal symmetry. However, the electric field has to be rotated away from the $z$ axis to open a gap in the spectrum. Let $\Theta, \Phi$ denote the angles of the electric field axis in a spherical coordinate system with the lattice in the equatorial plane.
Then, the dipole-dipole interaction can be expressed as
\begin{align}
    H^\text{dd}_{ij} &= \frac{\kappa}{|\vec{R}_{ij}|^3} \bigg[ f_0(\Theta, \phi_{ij} - \Phi)\, \Big( d^0_i d^0_j + \frac{1}{2}\big(d^+_i d^-_j + d^-_i d^+_j\big) \Big) \nonumber \\
                     &\hspace{1.38em} -\frac{3}{\sqrt{2}} \Big( f_1(\Theta, \phi_{ij} - \Phi) \, \big(d^0_i d^-_j + d^-_i d^0_j\big) + \hc \Big)\nonumber\\
                     &\hspace{2.2em} -\frac{3}{2} \big( f_2(\Theta, \phi_{ij} - \Phi) \, d^-_i d^-_j + \hc \big ) \bigg]
\end{align}
where
\begin{align}
    f_0(\Theta, \phi) &= 1-3\sin^2 \Theta \cos^2 \phi\, , \nonumber \\
    f_1(\Theta, \phi) &= \sin \Theta \cos \phi \bb{\cos \Theta \cos \phi + i \sin \phi} , \nonumber \\
    f_2(\Theta, \phi) &= \bb{\cos\Theta \cos\phi + i \sin \phi}^2 .
\end{align}
For $\Theta=0$, the interaction reduces to expression~(2) given in the main text.
For the tunneling rates we find
\begin{align}
    t^{++}_{ij} &= -\frac{\kappa d_{1}^2}{2R_{ij}^3} f_0(\Theta, \phi_{ij} - \Phi), \nonumber \\
    t^{11}_{ij} &= \frac{\kappa d_{0}^2}{R_{ij}^3} f_0(\Theta, \phi_{ij} - \Phi), \nonumber \\
    t^{1+}_{ij} &= \frac{3\kappa d_{0}d_{1}}{\sqrt{2}R_{ij}^3} f_1(\Theta, \phi_{ij} - \Phi)
\end{align}
where $d_0=\absv{\braketop{1}{d^0}{0}}$. Note that $t^{1+}_{ij}=0$ for $\Theta=0$, leading to the gapless spectrum for an electric field perpendicular to the lattice.

\section{Double-layer picture}

\begin{figure}[t]
    \centering
    \includegraphics[width=.9\columnwidth]{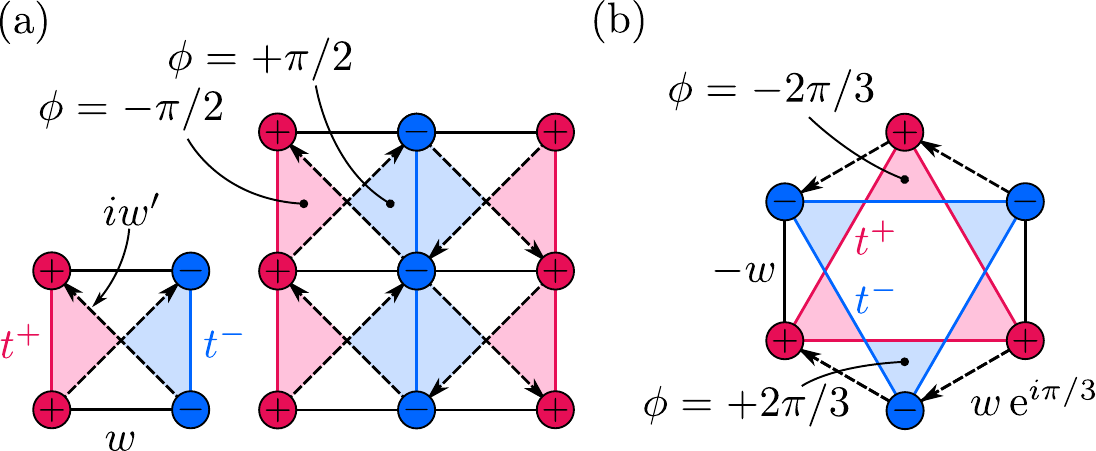}

    \caption{
        \label{fig:fig5}
        Hopping strengths and flux pattern of a single layer in different lattices.
        Tunneling elements without arrow are real numbers.
        Complex hoppings have the indicated strength along the arrow and the complex conjugate in the opposite direction.
        (a)~Square lattice: A single layer can be constructed by stripes of one component along one of the primitive vectors, effectively doubling the unit cell.
        The second layer is given by a translation along the second primitive vector.
        (b)~Honeycomb lattice: By distributing the $\ketp, \ketm$ orbitals to the two distinct sublattices it is possible to retain the symmetry of the lattice.
        The second layer is given by a $60^\circ$ rotation.
    }
\end{figure}

Topological band structures can be classified by considering equivalence classes of models that can be continuously deformed into each other without closing the energy gap~\cite{Hasan2010}.
In particular, the Chern number of a single band can only change if it touches another band.
Using this idea, we show that the model introduced in the main text in its $C=2$ phase is adiabatically equivalent to a system of two uncoupled copies of a $C=1$ layer.

To see this, imagine separating the two orbitals $\ketp$ and $\ketm$ per site spatially along the $z$-direction (without changing any tunneling rates) such that we obtain two separate square lattice layers, called A and B.
Sorting all terms in the Hamiltonian into intra- and inter-layer processes, we can write
\begin{align}
H=H_\text{A} + H_\text{B} + \lambda H_\text{AB}
\end{align}
where $\lambda=1$.
The choice which orbital resides in layer A (and B) can be made individually for each lattice site.
In any case, the resulting two layers will be interconnected by an infinite number of tunneling links $H_\text{AB}$.
The idea is to find a specific arrangement of the orbitals such that we can continuously let $\lambda \longrightarrow 0$ \emph{without} closing a gap in the excitation spectrum, preserving the topological phase while disentangling the layers.

Focusing on layer A (layer B being simply the complement), one possible arrangement is shown in~\figref{fig5}(a).
The ${+}$ (${-}$) orbitals are assigned to odd (even) columns along the $y$-direction.
For the Chern number of such a single layer we find $C=1$, using methods analogous to the ones described in the main text.
The full system can be understood as two such layers, shifted by one lattice site in $x$-direction.
With a unit cell twice the size of the original model, each layer contributes to one half of the full Brillouin zone, effectively doubling the Chern number to $C=2$.

The single layer system has some interesting properties.
In~\figref{fig5}(a) we show that it is possible to find a staggered magnetic flux pattern which creates the same tunneling phases as the dipole-dipole interaction, including tunneling up to the next-to-nearest neighbor level.
Using a site-dependent microwave dressing, it has been shown that a model similar to our single-layer system can be realized, giving rise to a $\nu=1/2$ fractional Chern insulating phase~\cite{Yao2012,Yao2013}.

On the honeycomb lattice, a single layer can be constructed which retains the original symmetry of the lattice, see~\figref{fig5}(b). Here, the two bands of the single layer also have $C=\pm 1$ but occupy the same Brillouin zone as the double layer system. Consequently, the four bands of the full system are constructed from the combination of two $C=1$ and two $C=-1$ bands, giving rise to a multitude of different topological phases.

\end{appendix}

\end{document}